\documentclass[english, pra, twocolumn]{revtex4}
\usepackage[T1]{fontenc}
\usepackage[latin9]{inputenc}
\usepackage{amsmath}
\usepackage{amssymb}

\usepackage{times}
\usepackage{babel}

\begin{document}

\title{Preparation of many-body states for quantum simulation}

\author{Nicholas J. Ward}
\affiliation{Department of Chemistry and Chemical Biology, Harvard University,
Cambridge, MA 02138}
\author{Ivan Kassal}
\affiliation{Department of Chemistry and Chemical Biology, Harvard University,
Cambridge, MA 02138}
\author{Al{\'a}n Aspuru-Guzik}
\email{aspuru@chemistry.harvard.edu}
\affiliation{Department of Chemistry and Chemical Biology, Harvard University,
Cambridge, MA 02138}

\begin{abstract}
While quantum computers are capable of simulating many quantum systems
efficiently, the simulation algorithms must begin with the
preparation of an appropriate initial state. We present a method for
generating physically relevant quantum states on a lattice in real
space. In particular, the present algorithm is able to prepare general
pure and mixed many-particle states of any number of particles. It
relies on a procedure for converting from a second-quantized state
to its first-quantized counterpart. The algorithm is efficient in
that it operates in time that is polynomial in all the essential
descriptors of the system, such the number of particles, the resolution
of the lattice, and the inverse of the maximum final error.
This scaling holds under the assumption that the wavefunction to be
prepared is bounded or its indefinite integral known and that the
Fock operator of the system is efficiently simulatable.
\end{abstract}

\maketitle

\section{Introduction}

Simulating quantum systems on a conventional computer requires
resources that generally scale exponentially with the size of the
system. Feynman proposed to solve this problem using a quantum machine
that would be able to mimic the properties of the quantum system \citep{feynman_simulating_1982}.
Subsequently, it has been demonstrated that quantum computers would
be able to simulate the time-dependent Schr{\"o}dinger equation for many
systems of interest using resources that scale polynomially with the
size of the system \citep{lloyd_universal_1996,christof_zalka_simulating_1998,wiesner_simulations_1996,lidar_calculatingthermal_1998,aspuru-guzik_simulated_2005,kais,kassal_quantum_2008,h2_experiment}.
However, all such simulations require the preparation of an appropriate
initial state, which must be preparable to within a chosen error.

In this work, we focus on the preparation of states on a gate-model
quantum computer. Our techniques can therefore complement those 
developed for the preparation of states in other models of quantum
computation, such as adiabatic quantum computing \citep{aharonov_adiabatic_2003,aspuru-guzik_simulated_2005}.

In general, we will call a state on $n$ qubits ``efficiently preparable''
if it can be prepared, to within error $\varepsilon$, using $\mathrm{poly}(n,\varepsilon^{-1})$
elementary (one- and two-qubit) quantum gates. Unfortunately, the
efficiently preparable states form only a small subset of all quantum states.
This is because a general state on $n$ qubits
contains $2^n$ amplitudes, and therefore one needs $O(2^n)$ gates to prepare it \citep{nielsen_quantum_2000}.
Indeed, state-preparation algorithms are known that almost reach this lower
bound \citep{mottonen_transformation_2005,bergholm_quantum_2005}.

In this work, we show that if wavefunctions are represented on a grid
in real space, then most quantum states of physical interest are
efficiently preparable. This is of interest because efficient,
grid-based simulation algorithms are known for physically realistic
systems \citep{christof_zalka_simulating_1998,wiesner_simulations_1996,lidar_calculatingthermal_1998,kassal_quantum_2008}.

Our work extends that of Zalka, who, in introducing real-space quantum
simulation \citep{christof_zalka_simulating_1998}, also provided the first
state preparation algorithm. However, his procedure is able to prepare
only states of single particles or uncorrelated many-particle systems.
We show how to use Zalka's single-particle wavefunctions
as building blocks, permitting the preparation of general
superpositions and mixed states of an arbitrary number of particles.
Our approach is motivated by electronic-structure theory, in that we
we choose particularly convenient single-particle bases in which
to expand more complicated states. We use the single-particle
eigenstates to form Slater determinants (configurations),
superpositions of which are used to express general many-particle
states.

Our scheme is essentially a method for translating states in
second quantization to the corresponding states in first quantization.
This has two advantages. First, many useful states that might be
needed in first quantization are easily prepared in second
quantization \citep{wang_efficient_2009}. In particular, we can prepare eigenstates
of operators if our scheme is combined with full configuration interaction
(FCI) \citep{szabo_ostlund}, an exact diagonalization method. FCI
is classically an exponentially hard problem due to the exponential
growth of the number of configurations with system size, but it can
be computed on a quantum computer in polynomial time \citep{aspuru-guzik_simulated_2005}.
The quantum FCI operates in second quantization, and can compute, for
example, the ground state wavefunction of a molecular system.
The second benefit of our method is that it is often easier to simulate
time-evolution in real space than in Fock space. For instance,
every simulation in second quantization would require
a separate set of basis-set--dependent operators and there might be
some processes, such as ionization, which could not be adequately
described using a small, localized basis set. In first-quantization,
however, all problems of chemical interest can be efficiently
simulated by direct simulation of the molecular Hamiltonian in
real space \citep{kassal_quantum_2008}.

This paper is organized as follows. We first consider the preparation
of many-particle states in which all the particles are the same. There
are three steps: the preparation of single-particle eigenstates in
a chosen basis, the preparation of many-particle configurations, and
finally the preparation of superpositions of configurations. We discuss
the preparation of mixed states, after which we turn to systems with
many different types of particles. Again, we consider the preparation
of configurations, their superpositions, and mixed states. We close
by showing that the algorithm is efficient in that its run-time is
polynomial in the size of the system, the number of qubits used to
encode the wavefunction, and the inverse of the maximum allowed error.

\section{One type of particle}

\label{sec:onetype}

Our algorithm translates from second to first quantization, and therefore
depends on the basis which is chosen for the representation of the
second-quantized states. We require a finite orthonormal basis of
functions $\{\phi_{1},\dots,\phi_{M}\}$, which are the eigenstates of
a known operator $\hat{F}$ on an $M$-dimensional, single-particle
Hilbert space. In our analogy with electronic-structure
theory, $\hat{F}$ would be the Fock operator for a single
particle \citep{szabo_ostlund}, and indeed we expect that the algorithm would
be at its most useful if $\hat{F}$ is chosen as the Fock operator
of an actual system. Although the form of $\hat{F}$ can be arbitrary,
subject to a few restrictions below, we will take advantage of the
analogy and refer to $\hat{F}$ as the \emph{Fock operator}. We will
also say, for example, that two eigenstates of $\hat{F}$ are \emph{degenerate}
if their \emph{energies} (the corresponding eigenvalues) are the same.

To ensure that the overall state-preparation algorithm scales efficiently,
we require that $\hat{F}$ can be efficiently simulated on a quantum computer,
i.e., that the simulation time scales polynomially with the size of the system.
More precisely, if there are $m$ particles occupying the $M$ orbitals and the 
simulation is done on a grid of $2^l$ sites (see below), then, for any $t$ and
any $\varepsilon>0$, there exist a unitary $\hat{U}$, composed of
$\mathrm{poly}(m,M,l,t,\varepsilon^{-1})$ elementary (one- and two-qubit) quantum gates, such that
$\left\lVert \hat{U}-e^{-i\hat{F}t} \right\rVert \le \varepsilon.$
Intuitively, this means that given an initial state, the final state generated
by the action of $\hat{F}$ for time $t$ can be calculated with reasonable effort
and reasonable error.

Several classes of Hamiltonians are known to be efficiently simulatable,
and together they ensure that most physically relevant Fock operators
will also be efficiently simulatable.
Very generally, an operator can be efficiently simulated if its matrix
in a given basis is sparse \citep{aharonov_adiabatic_2003,childs_thesis,berry_sparse}.
In particular, this includes Hamiltonians that are sums of local
operators, each of which acts on only a few degrees of freedom \citep{lloyd_universal_1996,nielsen_quantum_2000}.
In addition, many physically realistic real-space Hamiltonians
(such as those for chemical systems) can be efficiently
simulated \citep{christof_zalka_simulating_1998,wiesner_simulations_1996,lidar_calculatingthermal_1998,kassal_quantum_2008}.

We finally note that the requirement that the basis be orthonormal
may exclude certain commonly used basis sets. Many of the usually
encountered bases are appropriate,
such as plane waves or molecular orbitals, which diagonalize the molecular
Hartree-Fock Hamiltonian. However, non-orthogonal bases, such
as Gaussian wavepackets or atomic orbitals on more than one atomic
center, are not suitable for state preparation using our procedure.

\subsection{Single-particle eigenstates}

\label{sub:single}

A single-particle basis function $\phi$ can be prepared on a grid
by the state preparation method first proposed by Zalka \citep{christof_zalka_simulating_1998}
and rediscovered independently by both Grover and Rudolph \citep{grover_creating_2002}
and Kaye and Mosca \citep{kaye_quantum_2004}. The algorithm first prepares
the absolute value of the function, followed by
the addition of the phases. Specifically, given a register of $l$
qubits, representing a grid of $2^{l}$ points, and a basis state
$\phi(x)$ normalized over a length $L$, the algorithm first performs
the transformation\[
|0\rangle\rightarrow\left|\phi\right\rangle =\sum_{x=0}^{2^{l}-1}\left|\phi\left(x\frac{L}{2^{l}}\right)\right||x\rangle,\]
where each integer-valued state $|x\rangle$ is a position on the suitably
scaled grid. This state is generated from the state $\left|000\ldots\right\rangle $
by redistributing its amplitude $l$ times across the eigenstates $|x\rangle$.
To perform the redistribution correctly, we calculate the integrals\begin{equation}
I_{i,k}=\frac{\int_{k\frac{L}{2^{i}}}^{(k+1)\frac{L}{2^{i}}}|\phi(x)|^{2}dx}{\int_{k\frac{L}{2^{i}}}^{(k+2)\frac{L}{2^{i}}}|\phi(x)|^{2}dx},\label{eq:integrals}\end{equation}
for $k=0,\dots,2^{i}-2$ and $i=1,\dots,l$. The fraction $I_{i,k}$
is simply the probability that a particle in the $(k+1)$th subdivision
of size $L/2^{i}$ is also in its left half. If the denominator in
$I_{i,k}$ is zero, there is no amplitude to redistribute, so we can
skip this step. The first split is realized by performing a rotation
on the first qubit by $\arccos(\sqrt{I_{1,0}})$, corresponding to
the transformation\[
|0,\dots\rangle\rightarrow\sqrt{I_{1,0}}|0,\dots\rangle+\sqrt{1-I_{1,0}}|1,\dots\rangle.\]
This splits the norm of the initial state so that the appropriate
proportion is present on each half of the grid. The subsequent finer
splits are carried out in superposition using controlled rotations
on each qubit. For example, after the second iteration, the correct
proportion of the norm is present in each quarter of the grid. After
$l$ iterations, one obtains the desired state. Note that adding a
single qubit and the corresponding rotation doubles the precision
of the grid. Consequently, the absolute value of the wavefunction can be
efficiently approximated to any desired accuracy. Phases can be added
where necessary through phase-kickback \citep{cleve_quantum_1998}.
Given a procedure that can transform $|x\rangle\rightarrow e^{i\arg\phi(x)}|x\rangle$,
we can complete the preparation of $\left|\phi\right\rangle $ as\begin{eqnarray*}
\sum_{x=0}^{2^{l}-1}\left|\phi\left(x\frac{L}{2^{l}}\right)\right||x\rangle & \rightarrow & \sum_{x=0}^{2^{l}-1}e^{i\arg\phi\left(xL/2^{l}\right)}\left|\phi\left(x\frac{L}{2^{l}}\right)\right||x\rangle\\
 & = & \sum_{x=0}^{2^{l-1}}\phi\left(x\frac{L}{2^{l}}\right)|x\rangle=\left|\phi\right\rangle .\end{eqnarray*}

The same algorithm can be straightforwardly generalized to a three-dimensional
grid, where the position eigenstates are in Cartesian coordinates
and the corresponding three-dimensional integrals are used. In addition,
particle spin can be represented using additional qubits. A particle
with spin $S$ requires $\left\lceil \log_{2}(2S+1)\right\rceil $
qubits to store its $z$-projection $m_{S}$. In particular, there
is a natural mapping between the spin of spin-$\frac{1}{2}$ particles
and the states of a single qubit. If the Fock operator is spin-free,
the eigenstates will have separable spatial and spin degrees of freedom,
making the complete single-particle state $|\phi\rangle|m_{S}\rangle$.
Preparing the spin part of this wavefunction is relatively easy, for
it suffices to initialize the spin register in an integer state. If,
however, the eigenstate has correlation between the spatial and spin
degrees of freedom, it can be prepared using the techniques in Secs.
\ref{sub:superpositions} and \ref{sec:manytypes}. That is, we treat
the particle as if it were a composite system---composed of a spinless,
spatial part and a spin---and prepare its eigenstate using the techniques
below. In what follows, we will assume that our particles are fermions
and we will note where the algorithm would need to be modified for
bosons.

\subsection{Computational complexity of integration}

\label{sub:efficiency}

The preceding method for preparing single-particle states requires
the evaluation of integrals \eqref{eq:integrals}. Since this must
be performed in superposition, the integrals must be computed on the
quantum computer: precomputing them classically would require an exponentially
large look-up table. Consequently, the computational complexity of the state preparation procedure
will depend on the the cost of computing the integrals \citep{christof_zalka_simulating_1998,grover_creating_2002}. 

An integration procedure will, given a function $\phi:V\subset\mathbb{R}^{d}\rightarrow\mathbb{R}$
(where $V$ is a bounded region), supply an estimate $\tilde{I}$
of the integral $I=\int_{V}\phi(\mathbf{x})d\mathbf{x}$ such that
$|\tilde{I}-I|\leq \varepsilon_{I}$ with a certain fixed probability
$\delta$ (we'll call this condition the $(\varepsilon_{I},\delta)$
absolute error).

Integrals can be evaluated either analytically or numerically. If
the indefinite integrals of the basis functions are known, the definite
integrals over any box on the Cartesian grid can be computed. The
values of the indefinite integrals themselves can usually be computed
efficiently (i.e., with polynomial cost in the desired accuracy) because
they usually contain simple mathematical functions. In particular,
the time it takes to retrieve $n$ digits of any elementary function
is a polynomial in $n$ \citep{borwein}, and likewise for compositions
of elementary functions.

If the indefinite integrals are either unknown or impractical to compute,
numerical techniques can be used. In particular, any classical numerical
technique can, in principle, be implemented on a quantum computer.
For example, computing $\tilde{I}$ by Monte Carlo requires, in the
worst case \citep{fishman},\[
\left\lceil \left(\Phi^{-1}(1-\delta/2)\right)^{2}\sigma^{2}/\varepsilon_{I}^{2}\right\rceil \]
samples of $\phi$ for an $(\varepsilon_{I},\delta)$ absolute error,
where $\sigma^{2}$ is an estimate of the variance of $\phi$ over
$V$ and $\Phi(z)=(2\pi)^{-1/2}\int_{-\infty}^{z}\exp(-u^{2}/2)du$
is the standard normal cumulative distribution function. In particular,
if $\phi$ is bounded so that $\phi_{L}\le\phi(\mathbf{x})\le\phi_{U}$
for all $\mathbf{x}\in V$, the number of required samples is limited \citep{fishman} to\[
\left\lceil \left(\Phi^{-1}(1-\delta/2)(\phi_{U}-\phi_{L})/2\varepsilon_{I}\right)^{2}\right\rceil .\]
That is, Monte Carlo integration of any finite-variance function requires time that scales as $O(\varepsilon_{I}^{-2})$.
Acceptable wavefunctions need not be continuous or even finite \citep{peres}
(and hence may have infinite variance), but such examples are rather
contrived and rarely encountered in practice (but see below for $\delta$-functions).

Furthermore, it is known that quantum computers are able to offer
a quadratic speed-up over conventional probabilistic methods of integral
evaluation. Quantum integration techniques \citep{grover_framework_1998,abrams_fast_1999}
rely on amplitude amplification \citep{brassard_quantum_2000} to
achieve a computational complexity of $O(\varepsilon_{I}^{-1})$.
This has been proven optimal by Nayak and Wu \citep{nayak_quantum_1998,novak_quantum_2001}.
These techniques have the same general applicability as classical
Monte Carlo, and will likewise succeed for any bounded function. Furthermore,
the state preparation scheme of Soklakov and Schack \citep{soklakov_efficient_2006},
which relies on amplitude amplification, also succeeds in $O(\varepsilon_{I}^{-1})$
time. 

The preceding assumes that the function that we seek to prepare does
not depend substantially on the grid spacing. We would expect that
of realistic wavefunctions, assuming that the grid spacing is smaller
than the smallest wavelength of the system. A useful exception are
Kronecker $\delta$-functions, defined on a grid of $2^{l}$ points
as $\phi(\mathbf{x})=2^{l/2}\delta_{\mathbf{x},\mathbf{x}_{0}}$, where $\mathbf{x}_0$ is a
constant vector. The variance of $\delta$-functions grows exponentially
in $l$, and therefore they cannot be integrated efficiently by Monte
Carlo or prepared efficiently using the method of Soklakov and Schack \citep{soklakov_efficient_2006}. However, they can still be prepared
efficiently using our techniques because their indefinite integral,
the Heaviside function, can be easily computed in time independent
of $l$.

It remains to be shown that an error in the evaluated integral translates
to a comparable error in the prepared function. If the integrals \eqref{eq:integrals}
have a maximum error $\varepsilon_{I}$, that is, $\left|\tilde{I}_{i,k}-I_{i,k}\right|\leq \varepsilon_{I}$,
and the error in the final prepared state $\left|\tilde{\phi}\right\rangle $
is $\varepsilon_{\phi}=1-\left|\left\langle \left.\tilde{\phi}\right|\phi\right\rangle \right|$,
we find that $\varepsilon_{\phi}\leq l\varepsilon_{I}/2$. In the case
$l=1$, $\left|\phi\right\rangle =\sqrt{I}\left|0\right\rangle +\sqrt{1-I}\left|1\right\rangle $
and $\left|\tilde{\phi}\right\rangle =\sqrt{\tilde{I}}\left|0\right\rangle +\sqrt{1-\tilde{I}}\left|1\right\rangle$.
Then, assuming $0\le\tilde{I}\le1$, which is necessary for $\left|\tilde{\phi}\right\rangle $
to be an acceptable state,\begin{eqnarray*}
\varepsilon_{\phi} & = & 1-\sqrt{I\tilde{I}}-\sqrt{(1-I)(1-\tilde{I})}\\
 & \leq & 1-\sqrt{1-\varepsilon_{I}}\\
 & \approx & \frac{\varepsilon_{I}}{2},\end{eqnarray*}
where we have assumed that $\varepsilon_I\ll1$. For larger $l$, a
similar analysis applies qubit-wise: one finds that 
$\left\langle \left.\tilde{\phi}\right|\phi\right\rangle \geq \left(1-\varepsilon_{I}\right)^{l/2}\geq1-l\varepsilon_{I}/2$
(the last inequality holds for all $\varepsilon_I$ if $l\ge2$), whence
$\varepsilon_{\phi}\leq l\varepsilon_{I}/2$. That is to say, the error
in the prepared state grows only polynomially with the error in the
evaluated integral, a fact that we will use later on to establish
the computational cost of the state preparation algorithm.

\subsection{Many-particle eigenstates}

\label{sub:many}

The next step is to use Zalka's algorithm to prepare multi-particle
configurations. That is, we wish to prepare the position-space representation
of a second-quantization state $\left|n_{1}n_{2}\dots n_{M}\right\rangle _{2nd}$
(a Fock eigenstate), where $n_{i}$ is the occupation number of the
basis orbital $\phi_{i}$ $(1\le i\le M)$. The position-space representation
of $\left|n_{1}n_{2}\dots n_{M}\right\rangle _{2nd}$ will be a Slater
determinant of the occupied orbitals, and it will be an eigenstate
of the many-body \emph{Hartree-Fock Hamiltonian}\begin{equation}
\hat{H}=\sum_{i=1}^{m}\hat{F}_{i},\label{eq:HF}\end{equation}
where $m$ is the total number of particles and $\hat{F}_{i}$ is
the Fock operator $\hat{F}$ acting on the particle $i$ \citep{szabo_ostlund}.

We assume that the state $\left|n_{1}n_{2}\dots n_{M}\right\rangle _{2nd}$
has already been prepared by some previous algorithm. The $M$ basis
orbitals are occupied by $m$ particles and we let $j_{1},\dots,j_{m}$
be the indices of the occupied orbitals. We therefore wish to perform
the transformation
\begin{multline}
|n_{1}n_{2}\dots n_{M}\rangle_{\mbox{\tiny$2nd$}}\otimes|0\dots\rangle_{\mbox{\tiny$1st$}} \rightarrow \\
|n_{1}n_{2}\dots n_{M}\rangle_{\mbox{\tiny$2nd$}}\otimes\frac{1}{\sqrt{m!}}\sum_{\sigma\in S_{m}}\mathrm{sgn}(\sigma)|\sigma(\phi_{j_{1}}\phi_{j_{2}}\dots\phi_{j_{m}})\rangle_{\mbox{\tiny$1st$}} \\
= |n_{1}n_{2}\dots n_{M}\rangle_{\mbox{\tiny$2nd$}}\otimes\left|\Phi\right\rangle_{\mbox{\tiny$1st$}}, \label{eq:many}
\end{multline}
which takes the input state and prepares the appropriate first-quantized
Slater determinant $\left|\Phi\right\rangle _{1st}$, a
superposition of all the permutations on the $m$ occupied orbitals
($S_{m}$ is the symmetric group on $m$ elements and sgn denotes
the signature). Here $|0\dots\rangle_{1st}$ contains $m$ registers
for the $m$ first-quantized occupied orbitals
$\left|\phi_{j_{1}}\right\rangle_{1st} ,\dots,\left|\phi_{j_{m}}\right\rangle_{1st}$.
Note that \eqref{eq:many} is not in general a reversible operation,
as multiple input states would be mapped to the same antisymmetrized
result. To ensure the algorithm is reversible, we additionally require \citep{abrams_simulation_1997}
that $j_{1}<j_{2}<\dots<j_{m}$. The procedure can be slightly modified if
bosons are in question: then $\mathrm{sgn}(\sigma)$ is to be omitted,
and the $j_{i}$ must satisfy $j_{1}\le j_{2}\le\ldots\le j_{m}$.

The transformation \eqref{eq:many} is accomplished in two steps. First, the
occupied single-particle basis orbitals are each prepared in a
separate register, forming a Hartree product:
\begin{multline*}
|n_{1}n_{2}\dots n_{M}\rangle_{2nd}\otimes|0\dots\rangle_{1st}\rightarrow \\
|n_{1}n_{2}\dots n_{M}\rangle_{2nd}\otimes|\phi_{j_{1}}\phi_{j_{2}}\dots\phi_{j_{m}}\rangle_{1st}.
\end{multline*}
The procedure can be modified in the case of bosons by counting the
occupation of each orbital and preparing that many copies in separate
registers.

In the next step, the Hartree product is antisymmetrized,
which produces the desired Slater determinant.
To complete this step, we introduce an improved form of the
antisymmetrization algorithm developed by Abrams and Lloyd \citep{abrams_simulation_1997}.
The algorithm begins with the $m$ wavefunctions $\left|\phi_{j_{i}}\right\rangle $
to be antisymmetrized in register $A$, and $m \left\lceil \log_{2}m\right\rceil $
qubits in register $B$ (where each grouping of $\left\lceil \log_{2}m \right\rceil$ qubits
constitutes a ``quword'') initialized to $|0\rangle$. Using a
series of controlled rotations, $B$ is converted to the state\[
\frac{1}{\sqrt{m!}}\sum_{i=1}^{m}|i\rangle_{B[1]}\otimes\sum_{i=1}^{m-1}|i\rangle_{B[2]}\otimes\cdots\otimes|1\rangle_{B[m]},\]
which is a superposition of $m!$ unique states consisting of $m$
quwords each, and $B[i]$ denotes the $i$th quword in register $B$.
Next we will transform this state into the superposition\[
\frac{1}{\sqrt{m!}}\sum_{\sigma\in S_{m}}|\sigma(1,\dots,m)\rangle_{B}\]
as follows. First let $B^{\prime}[1]=B[1]$. Then assign to
$B^{\prime}[i]$ the $B[i]$th natural number not present
in the set $\{B'[1],B'[2],\dots,B'[i-1]\}$. This leaves the quantum
computer in the state\begin{equation}
\frac{1}{\sqrt{m!}}|\phi_{j_{1}}\phi_{j_{2}}\dots\phi_{j_{m}}\rangle_{A}\otimes\sum_{\sigma\in S_{m}}|\sigma(1,\dots,m)\rangle_{B}.\label{eq:sort_before}\end{equation}
Register $B$ now contains a symmetrized state and this symmetry can
be transferred to register $A$ by sorting $B$ while performing the
same swaps on the wavefunctions in $A$. This yields the symmetrized
state\begin{equation}
\frac{1}{\sqrt{m!}}\sum_{\sigma\in S_{m}}|\sigma(\phi_{j_{1}}\phi_{j_{2}}\dots\phi_{j_{m}})\rangle_{A}\otimes|1,2,\ldots,m\rangle_{B},\label{eq:sort_after}\end{equation}
which is what we would keep if we were interested in preparing bosonic
states. To instead obtain an antisymmetrized state, we need only count
the number of exchanges made in the sort, and reverse the sign
of the wavefunction if
it is odd. If we now eliminate the register $B$, $A$ contains the
desired multi-particle state $|\Phi\rangle_{1st}$.

The original algorithm, introduced by Abrams and Lloyd, included an
additional auxilliary register $C$, which would then be used as an
intermediate for the sorting of $A$ and $B$. We eliminate this step
by sorting $A$ and $B$ together directly.

\subsection{Superpositions}

\label{sub:superpositions}

We now generalize the algorithm to the preparation of superpositions
of many-particle states. Given a superposition of second-quantization
states $|\mathbf{n}_{i}\rangle_{2nd}=|n_{1i}n_{2i}\dots n_{Mi}\rangle_{2nd}$,
with amplitudes $\alpha_{i}$, we wish to perform the transformation\[\left(
\sum_{i}\alpha_{i}|\mathbf{n}_{i}\rangle_{2nd}\right)\otimes|0\dots\rangle_{1st}\rightarrow|0\dots\rangle_{2nd}\otimes\left(\sum_{i}\alpha_{i}|\Phi_{i}\rangle_{1st}\right).\]
The superposition on the left might come from a variety of sources.
For example, an easily-prepared equal superposition of Fock states
would result in an equal superposition of real-space wavefunctions.
Wang et al. provide an algorithm for preparing general superpositions
of Fock states on a quantum computer \citep{wang_efficient_2009}.
Alternatively, a quantum electronic-structure algorithm could be used
to efficiently produce a physically relevant superposition.
For example, an FCI algorithm could specify the ground state
of a chemical or other many-body system in terms of a superposition
of Fock states \citep{aspuru-guzik_simulated_2005}.

As before, we begin by applying Zalka's state preparation algorithm
to the input state. Because this linear operation is carried
out in superposition, it accomplishes the transformation
\begin{multline*}\left(
\sum_{i}\alpha_{i}|\mathbf{n}_{i}\rangle_{2nd}\right)\otimes|0\dots\rangle_{1st}\rightarrow \\
\sum_{i}\alpha_{i}|\mathbf{n}_{i}\rangle_{2nd}\otimes|\phi_{j_{1}i}\phi_{j_{2}i}\dots\phi_{j_{m}i}\rangle_{1st}.
\end{multline*}
Note that the single-particle wavefunctions are now entangled with
the input state. For a multi-particle eigenstate, the situation was
different because the resulting state was separable. Hence, to
separate the first-quantized wavefunctions from the second-quantized
ones, we must ``uncompute'' the second-quantized states.
This must be accomplished using only manipulations on the
register containing the first-quantized wavefunctions $|\phi_{i}\rangle_{1st}$:
if we can regenerate the input state from the wavefunctions,
the input register can be reset to $|0\rangle_{2nd}$ as desired.
Given the one-to-one correspondence between a second quantization
state and the corresponding first quantization wavefunctions, regenerating
the input state amounts to the problem of identifying the wavefunctions
$|\phi_{i}\rangle_{1st}$ given only the information contained in their
first-quantized representation.

For non-degenerate eigenstates, each state $|\phi_{i}\rangle_{1st}$
can be uniquely identified using its energy, which can be obtained
through the phase estimation procedure \citep{kitaev_quantum_1995,cleve_quantum_1998,abrams_quantum_1999}.
In general, given a unitary $\hat{U}$ and its eigenstate $|\psi\rangle$,
the phase estimation algorithm finds the eigenvalue of $|\psi\rangle$.
Specifically, since $\hat{U}|\psi\rangle=e^{2\pi i\theta}|\psi\rangle$,
we have $\hat{U}^{k}|\psi\rangle=e^{2\pi ik\theta}|\psi\rangle.$
By controlled applications of the powers of $\hat{U}$ to $|\psi\rangle$,
controlled on the state $\frac{1}{2^{q/2}}\sum_{k=0}^{2^{q}-1}\left|k\right\rangle $,
one gets $\frac{1}{2^{q/2}}\sum_{k=0}^{2^{q}-1}\left|k\right\rangle \hat{U}^{k}\left|\psi\right\rangle =\frac{1}{2^{q/2}}\sum_{k=0}^{2^{q}-1}e^{2\pi ik\theta}\left|k\right\rangle \left|\psi\right\rangle $.
An efficient quantum Fourier transform on the control qubits will
now yield the first $q$ digits of the binary expansion of $\theta$.
If we choose $\hat{U}$ such that $\hat{U}|\phi_{i}\rangle_{1st}=e^{2\pi iE_{i}}|\phi_{i}\rangle_{1st}$,
we can use phase estimation with enough control qubits to obtain
an approximation of the energies $E_{i}$. In particular, the natural choice
$\hat{U}=e^{-i\hat{H}t}$, where $\hat{H}$ is the Hartree-Fock Hamiltonian
\eqref{eq:HF}, supplies the appropriate unitary for a suitable choice of
the time $t$. Note that $\hat{U}$ can be simulated efficiently
because $\hat{H}$ is a sum of Fock operators which are efficiently simulatable
by assumption. The energy eigenvalues are stored in an additional
register containing enough qubits to provide precision that distinguishes
between nearby energies. 

In the case that the spectrum of $\hat{H}$ is degenerate, properties
other than the energy of the states need to be used to distinguish
them. If the degeneracy is caused by a symmetry of the Hamiltonian,
the elements of the symmetry group can be used for this discrimination,
as we outline in Sec. \ref{sub:structural}. If the degeneracies are
accidental, other techniques are required, and we give some suggestions
in Sec. \ref{sub:accidental}. In addition, the techniques in Sec.
\ref{sub:accidental} can be used for distinguishing eigenvalues that
are exponentially close together and therefore cannot be distinguished
efficiently by phase estimation.

Phase estimation using both $\hat{U}$ to find energy eigenvalues
and appropriate symmetry operations to distinguish degenerate states
will provide us with a unique combination of eigenvalues for each
state in the superposition. These eigenvalues can then be used (for
example in conjunction with a look-up table) to uniquely identify
the wavefunction and subtract 1 from the corresponding occupation
number vector of the second-quantization state. Because this is done
in superposition for every single-particle wavefunction, the input
state is converted to $|0\rangle_{2nd}$. \newline This accomplishes the total
transformation\begin{multline}
 \left(\sum_{i}\alpha_{i}|\mathbf{n}_{i}\rangle_{2nd}\right)\otimes|0\dots\rangle_{1st}\rightarrow \\
 |0\dots\rangle_{2nd}\otimes\left(\sum_{i}\alpha_{i}|\phi_{j_{1}i}\phi_{j_{2}i}\dots\phi_{j_{m}i}\rangle_{1st}\right),\label{eq:superpositions}\end{multline}
which is a separable state. The antisymmetrization step can now proceed
in superposition as usual, resulting in the final state $|\Psi\rangle_{1st}=\sum_{i}\alpha_{i}|\Phi_{i}\rangle_{1st}$,
as desired. This completes the state-preparation algorithm for a given
superposition of multi-particle states.

\subsection{Resolving degeneracies caused by symmetry}

\label{sub:structural}

The procedure in Sec. \ref{sub:superpositions} assumes that it is
possible to distinguish eigenstates based on their energy. If there
are degenerate states, additional operations are required to distinguish
them. Degeneracies in quantum states usually arise as a result of
symmetry---degeneracies that do not are called {}``accidental''
and we treat them separately in Sec. \ref{sub:accidental}. For symmetry-caused
degeneracy, distinguishing degenerate states requires an understanding
of how they transform under the symmetry operations of the system.
All of the wavefunctions $|\phi_{i}\rangle_{1st}$ are eigenstates of each
symmetry operation within the point group, but degenerate wavefunctions
will always have different eigenvalues for at least one of the operations.
Phase estimation can still be used to obtain a unique set of eigenvalues,
but in addition to finding the energies, we can distinguish the wavefunctions
by symmetry. By applying phase estimation using an appropriate symmetry
operation as the unitary operator, we obtain additional eigenvalues
to distinguish degenerate states.

Because there are only a limited number of symmetries that are possible
in physical systems, it will rarely be necessary to use more than a few
readout qubits to retrieve all the distinguishing eigenvalues. With
the exception of systems with spherical, cubic, or icosahedral symmetry,
which we treat below, all systems in three-dimensional space have
a symmetry point group all of whose irreducible representations are
one- or two-dimensional \citep{cotton_chemical_1990}. Wavefunctions
transforming as the one-dimensional irreducible representations are
non-degenerate, while the ones transforming as the two-dimensional
irreducible representations come in degenerate pairs. Distinguishing
them, therefore, requires the determination of only one symmetry eigenvalue
which is different for the two wavefunctions. 

This is most easily done in the case of point groups $C_{nv}$, $C_{\infty v}$,
$D_{n}$, $D_{nh}$, $D_{\infty h}$, and $D_{nd}$, all of which
contain a $C_{2}$ axis or a reflection plane that has character zero
in all of the two-dimensional irreducible representations. In this
case, one of the two degenerate wavefunctions is invariant under the
reflection or $C_{2}$ rotation, while the other acquires a phase
of $-1$. To distinguish them, one would use the reflection or the
$C_{2}$ rotation as the unitary of phase estimation with one readout
qubit (note that these operations are easy to implement, being simple
linear transformations). The readout qubit, initialized in the state
$\left(\left|0\right\rangle +\left|1\right\rangle \right)/\sqrt{2}$,
would, under the action of the symmetry operation, be converted to
$\left(\left|0\right\rangle \pm\left|1\right\rangle \right)/\sqrt{2}$,
depending on the acquired phase. A Hadamard gate would then return
$\left|0\right\rangle $ or $\left|1\right\rangle $, perfectly discriminating
between the two eigenfunctions.

Symmetry groups $C_{n}$, $C_{nh}$, and $S_{2n}$ have, strictly
speaking, only one-dimensional irreducible representations. However,
there are pairs of representations that are complex conjugates of
each other, meaning that the corresponding energy levels are degenerate
due to time-reversal symmetry. These pairs of conjugate representations
are called ``separably degenerate \citep{bunker_molecular_2006}," and
the corresponding wavefunctions can be distinguished using the principal
symmetry axis $C_{n}$ (or $S_{2n}$ in the $S_{2n}$ groups). In
each case, under the action of $C_{n}$, one of the wavefunctions
acquires a phase $\omega$ and the other $\omega^{*}$, where $\omega=e^{2\pi i/n}$
(there are also cases where the pairs acquire phases such as $-\omega$
and $-\omega^{*}$, $\omega^{2}$ and $\left(\omega^{2}\right)^{*}$,
and so on, but these do not change the procedure outlined here). Phase
estimation can, as usual, measure this phase up to a certain precision.
However, since $1/n$ usually does not have a finite binary expansion,
there will be an associated error in the phase estimation. This can
be reduced below an arbitrary threshold by the addition of more readout
qubits, as discussed in Sec. \ref{sec:errors_and_scaling}. This is
especially true since real physical systems almost never have $C_{n}$
axes with $n>8$, meaning that only several qubits will be required
for readout.

The cubic and icosahedral groups, $T$, $T_{h}$, $T_{d}$, $O$,
$O_{h}$, $I$, and $I_{h}$, all have three-dimensional irreducible
representations (and $I$ and $I_{h}$ also have four- and five-dimensional
ones). Fortunately, there are plenty of reflection planes and $C_{2}$
axes which can be used for discrimination just as was done in the
simpler groups above. Distinguishing three or four degenerate states
requires two symmetry eigenvalue comparisons (and three in the case
of five-fold degeneracy). Consequently, two readout qubits are required
in these cases, one for each comparison (or three qubits in the five-fold
degenerate case).

Degenerate states of spherically symmetric systems, such as atoms,
can be distinguished by energy and by their angular momentum quantum
numbers $\ell$ and $m_{\ell}$. The maximally symmetric case is the
$1/r$ potential, where the conservation of the Laplace-Runge-Lenz
vector implies that all states with equal principal quantum number
$n$ are degenerate. If our basis contains states with $n\le n_{\mathrm{max}}$,
we would require $O(\log_{2}n_{\mathrm{max}})$ qubits for the
discrimination of the angular momentum states (that is, $O(\log_{2}n_{\mathrm{max}})$
qubits each for $\ell$ and $m_{\ell}$). While circumstances where
one encounters states of extremely high angular momentum are rare,
we can see that the discrimination can be performed efficiently. The
phase estimation in this case would use discrete rotations as its
unitary operator. A similar approach was suggested by Zalka for the
related problem of implementing unitary representations of
SU(2) \citep{zalka_implementing_2004}.

\subsection{Resolving accidental degeneracies and exponentially close eigenstates}

\label{sub:accidental}

In Sec. \ref{sub:structural}, we outlined a procedure for distinguishing
states that are degenerate because of symmetry. However, the eigenstates
might also be accidentally degenerate or exponentially close in energy
so that they cannot be efficiently distinguished by phase estimation.
In those cases, it is not possible to distinguish between the (near-)degenerate
states using the symmetry-based procedure. 

One way around these problems is to transform to another basis where
the (near-)degeneracy does not arise. A way of accomplishing this
is to use a perturbed Fock operator $\hat{F}^{\prime}=\hat{F}+\hat{V}$,
where $\hat{V}$ is a small, efficiently simulatable perturbation that
breaks the (near-)degeneracies. In a finite basis, $\hat{V}$ must
also be small to ensure that the new basis can adequately describe
the target state. The new eigenfunctions are obtained from the old
using perturbation theory, as are the new coefficients of the state
that we wish to prepare. This change of basis can be done efficiently
on a classical computer, before proceeding as normal with the state
preparation algorithm. For the purposes of phase estimation, the new
Fock operator can be efficiently simulated by operator splitting because
both $\hat{F}$ and $\hat{V}$ are efficiently simulatable.

A drawback of this procedure is that the perturbation may destroy
certain desirable symmetries of the system. In some cases, this can
be avoided if we choose $\hat{V}\propto\left|\phi_{i}\right\rangle \left\langle \phi_{i}\right|$,
where $\left|\phi_{i}\right\rangle $ is one of the (near-)degenerate
eigenstates. In that case, $\hat{F}'$ and $\hat{F}$ would have the
same eigenstates and no change of basis would be needed. Of course,
it is possible that $\hat{V}$ in this form is not efficiently simulatable, in which
case this scheme would not be efficient.

\subsection{Mixed states}

\label{sub:mixed}

The previous sections outline the procedure for preparing general
pure states, which in the chosen basis read\begin{equation}
\left|\Psi\right\rangle _{1st}=\sum_{i}\alpha_{i}|\Phi_{i}\rangle_{1st}.\label{eq:mixed1}\end{equation}
From now on, we drop the subscript $1st$ for clarity. We now wish
to prepare a mixed state with density operator \[
\hat{\rho}=\sum_{i}p_{i}\left|\Psi_{i}\right\rangle \left\langle \Psi_{i}\right|,\]
where $\left|\Psi_{i}\right\rangle $ are arbitrary pure states of
the form \eqref{eq:mixed1} and the probabilities $p_{i}$ add up
to 1. This scheme could be used for the preparation of thermal states,
in which case one would choose $\left|\Psi_{i}\right\rangle $ to
be the Hamiltonian eigenstates and $p_{i}=e^{-\beta E_{i}}/Z$, where
$\beta=1/k_{\mathrm{B}}T$ and $Z$ is the partition function.
Our approach to the thermalization problem is therefore different
from that of Terhal and DiVincenzo, who prepare thermal states by
simulating an external bath \citep{terhal_problem_2000}.

We assume that each $\left|\Psi_{i}\right\rangle $ can be efficiently
specified using some specification $\left|\xi_{i}\right\rangle $
(for example, $\left|\Psi_{i}\right\rangle $ is the $\xi_{i}$th
eigenstate of the Hamiltonian). We begin by preparing the state $\sum_{i}\sqrt{p_{i}}\left|\xi_{i}\right\rangle $.
This can be done using the procedure in Sec. \ref{sub:single}
if we order the $\xi_{i}$'s so that they may be thought of as a function
on a one-dimensional grid. We then run the entire state-preparation
algorithm in superposition, preparing the appropriate $\left|\Psi_{i}\right\rangle $
conditional on the value of the $\left|\xi_{i}\right\rangle $. This
yields the state\[
\left|\Xi\right\rangle =\sum_{i}\sqrt{p_{i}}\left|\xi_{i}\right\rangle \left|\Psi_{i}\right\rangle ,\]
the density operator of which is\[
\hat{\rho}_{\Xi}=\sum_{i,i'}\sqrt{p_{i}p_{i'}}\left|\xi_{i}\right\rangle \left\langle \xi_{i'}\right|\otimes\left|\Psi_{i}\right\rangle \left\langle \Psi_{i'}\right|.\]
Tracing out the specification register, we get the desired density
operator\begin{eqnarray*}
\hat{\rho} & = & \mathrm{Tr}_{\xi}\hat{\rho}_{\Xi}\\
 & = & \sum_{i,i'}\sqrt{p_{i}p_{i'}}\left|\Psi_{i}\right\rangle \left\langle \Psi_{i'}\right|\mathrm{Tr}\left|\xi_{i}\right\rangle \left\langle \xi_{i'}\right|\\
 & = & \sum_{i}p_{i}\left|\Psi_{i}\right\rangle \left\langle \Psi_{i}\right|.\end{eqnarray*}
In practical terms, tracing out the specification register amounts
to doing nothing at all. That is, each $\left|\Psi_{i}\right\rangle $
is entangled to a different $\left|\xi_{i}\right\rangle $, meaning
that the $\left|\Psi_{i}\right\rangle $'s evolve separately under
time evolution, as they would if they were independent members
of an ensemble.

One can notice that density operators diagonal in the $\left|\Phi_{i}\right\rangle $
basis can be prepared more directly. In the previous Sec. \ref{sub:superpositions}, we had
to ``disentangle'' the first- and second-quantized states. If
we had instead simply traced out the input register, we would have
obtained a mixed state diagonal in the $\left|\Phi_{i}\right\rangle $
basis.

\section{Many types of particles}

\label{sec:manytypes}

In Sec. \ref{sec:onetype}, we outlined an algorithm for the preparation
of arbitrary many-particle states (pure or mixed) of a system of identical
particles. However, one often wants to consider systems of more than
one type of particle, or treat particles of the same kind, but separated
in space, as different (the latter approach might be useful, for example,
in computing electron transfer matrix elements for large molecules) \citep{e_transfer}.
We consider the case of two types of particles,
with the generalization to more types being clear.

One wants to prepare an arbitrary two-particle state\[
\left|\Theta\right\rangle =\sum_{i,j}\alpha_{i,j}\left|\Phi_{A,i}\right\rangle \left|\Phi_{B,j}\right\rangle ,\]
where $\left|\Phi_{A,i}\right\rangle $ is a many-particle eigenstate
of particles of type $A$, and $\left|\Phi_{B,j}\right\rangle $ is
an eigenstate of particles of type $B$. Each element $\left|\Phi_{A,i}\right\rangle \left|\Phi_{B,j}\right\rangle $of
this superposition is easily created by preparing the appropriate
state in separate registers as was done in Sec. \ref{sub:many}. Creating
$\left|\Theta\right\rangle $ itself can be done in analogy to the
preparation of superpositions in Sec.\ref{sub:superpositions}. We
start by efficiently specifying $\left|\Theta\right\rangle $ using
occupation number vectors of the $\left|\Phi_{A,i}\right\rangle $
and the $\left|\Phi_{B,i}\right\rangle $, namely\[
\sum_{i,j}\alpha_{i,j}|\mathbf{n}_{A,i}\rangle_{2nd}|\mathbf{n}_{B,j}\rangle_{2nd}\otimes|0_{A}\rangle_{1st}|0_{B}\rangle_{1st}.\]
We then complete the state preparation, in superposition, as we did
in Sec. \ref{sub:superpositions}, treating each register separately.
Doing so produces $\left|\Theta\right\rangle $.

There are many circumstances in which the ability to prepare states
such as these would be valuable. For instance, in chemical dynamics
it is necessary to treat the nuclei and the electrons separately.
If we restricted our state preparation to simple product states such
as $\left|\Phi_{A,i}\right\rangle \left|\Phi_{B,j}\right\rangle $,
we would get a state in the Born-Oppenheimer approximation, which
is often a good approximation to the initial states of reactants participating
in chemical reactions. However, as the procedure for preparing $\left|\Theta\right\rangle $
shows, quantum computers could just as easily prepare non--Born-Oppenheimer
states in which there is correlation between electronic and nuclear
degrees of freedom.

Many-particle mixed states can likewise be prepared by following the
procedure in Sec. \ref{sub:mixed} separately for each type of particle.

\section{Errors and the computational cost}

\label{sec:errors_and_scaling}

For the state preparation algorithm to be considered efficient, the
time it takes to execute it must scale as a polynomial in the sizes
of the input. More precisely, it should scale as a polynomial in $l$,
the number of qubits used to store the wavefunction and $m$, the
number of occupied single-particle orbitals, which is the best descriptor
of the total size of the system.

In this section, we first show that pre-existing errors are amplified
at most linearly by subsequent steps of the algorithm. We then use
this fact to obtain the total computational cost of preparing an arbitrary
quantum state.

\subsection{Errors}

\label{sub:errors}

Assuming that the quantum gates are executed perfectly---or that the
gate errors are corrected using efficient error correction algorithms---there are
five sources of error in the state preparation algorithm:

\textbf{1. Preparation of single-particle eigenstates.} Zalka's method
that we adopt in Sec. \ref{sub:single} requires evaluation of the integrals
\eqref{eq:integrals}. We have addressed
the computational cost of integral evaluation in Sec. \ref{sub:efficiency},
where we show that the procedure can be accomplished in time polynomial
in $\varepsilon_{I}^{-1}$ if the wavefunction's indefinite integral
is known or, more generally, if the wavefunction is bounded. The resulting
error in the prepared single-particle eigenstate is $\varepsilon_{\phi}\leq l\varepsilon_{I}/2$.

\textbf{2. Assembly of many-particle eigenstates.} Many-particle eigenstates
\eqref{eq:many} inherit the errors present in the single-particle
eigenstates $\left|\phi_{i}\right\rangle $ that are used to assemble
them. Supposing that the prepared states $\left|\tilde{\phi}_{i}\right\rangle $
approximate the true states $\left|\phi_{i}\right\rangle $ with error
$\varepsilon_{\phi,i}=1-\left|\left\langle \left.\tilde{\phi}_{i}\right|\phi_{i}\right\rangle \right|$,
then the prepared Hartree product $\left|\tilde{\phi}_{j_{1}}\dots\tilde{\phi}_{j_{m}}\right\rangle $
suffers an error \begin{eqnarray*}
\varepsilon_{\Phi} & = & 1-\left|\left\langle \left.\tilde{\phi}_{j_{1}}\dots\tilde{\phi}_{j_{m}}\right|\phi_{j_{1}}\dots\phi_{j_{m}}\right\rangle \right|\\
 & = & 1-\prod_{i=1}^{m}(1-\varepsilon_{\phi,j_{i}})\\
 & \le & \sum_{i=1}^{m}\varepsilon_{\phi,j_{i}}\le m\varepsilon_{\phi}\leq ml\varepsilon_{I}/2,\end{eqnarray*}
where $\varepsilon_{\phi}=\max\varepsilon_{\phi,i}$. Since the total
error grows as a polynomial in both the single-state error and the
number of occupied states, the assembly of Hartree products amplifies the pre-existing errors only linearly in $m$.
The remaining step, the antisymmetrization of the Hartree product,
does not introduce additional errors.

\textbf{3. Preparation of superpositions.} The parallel state-preparation
that is used to perform the transformation \eqref{eq:superpositions}
does not introduce any additional errors with the exception of the
possible failures of phase estimation, discussed below. Nevertheless,
we should see how pre-existing errors propagate through this step.
If the prepared state is $\left|\tilde{\Psi}\right\rangle =\sum_{i}\alpha_{i}\left|\tilde{\Phi}_{i}\right\rangle $,
we see that it suffers an error with respect to the target state\begin{eqnarray*}
\varepsilon_{\Psi} & = & 1-\sum_{i}|\alpha_{i}|^{2}\left|\left\langle \left.\tilde{\Phi}_{i}\right|\Phi_{i}\right\rangle \right|\\
 & = & 1-\sum_{i}|\alpha_{i}|^{2}(1-\varepsilon_{\Phi,i})\\
 & \le & \varepsilon_{\Phi}\leq ml\varepsilon_{I}/2,\end{eqnarray*}
where $\varepsilon_{\Phi}=\mathrm{max}\varepsilon_{\Phi,i}$ and where
we have assumed that $\left\langle \left.\tilde{\Phi}_{i}\right|\Phi_{j}\right\rangle =0$
for $i\ne j$. In other words, the error in $\left|\tilde{\Psi}\right\rangle $
is limited by the error of its components.

\textbf{4. Discrimination of states in a superposition.} The preparation
of superpositions described in Secs. \ref{sub:superpositions}-\ref{sub:accidental}
and \ref{sec:manytypes} relies on phase estimation as a means of
distinguishing states. Since the eigenenergies will rarely have finite
binary expansions, there will be errors introduced at this step. If
two phases differ at the $n$th bit and we perform phase estimation
with $q=n+p$ qubits, the probability of an incorrect identification
is $1/2(2^{p}-2)$, meaning that the success probability will be $1-\varepsilon_{\mathrm{PE}}$
provided we implement phase estimation with $p=\left\lceil \log\left(2+1/2\varepsilon_{\mathrm{PE}}\right)\right\rceil$
additional qubits \citep{nielsen_quantum_2000}. The additional overhead,
logarithmic in $\varepsilon_{\mathrm{PE}}^{-1}$, does not compromise
the efficiency. The same arguments apply to the phase estimation of
eigenvalues of $C_{n}$ belonging to states in separably degenerate
irreducible representations of groups $C_{n}$, $C_{nh}$, and $S_{2n}$
(see Sec. \ref{sub:structural}). The symmetry eigenvalues that are
useful for states in the other point groups are always $\pm1$, and
can be perfectly resolved using phase estimation with a single readout
qubit.

In addition, failures of state discrimination based on
phase estimation can be detected after the state preparation is complete.
The second-quantized register, which should be uncomputed during
the procedure, should be measured at the end. If $\left|0\dots\right\rangle $
is observed, phase estimation will have succeeded. Otherwise, a
misidentification will have occurred, and the procedure ought to be repeated.
This simple, classical error correction introduces only a constant
overhead and ensures that phase estimation
does not contribute to the error in the final prepared state.

\textbf{5. Assembly of mixed states.} In the notation of Sec. \ref{sub:mixed},
if the prepared states $\left|\tilde{\Psi}_{i}\right\rangle $ approximate
the true states $\left|\Psi_{i}\right\rangle $ with an error $\varepsilon_{\Psi,i}=1-\left|\left\langle \left.\tilde{\Psi}_{i}\right|\Psi_{i}\right\rangle \right|$,
and assuming perfect preparation of the state $\sum_{i}\sqrt{p_{i}}\left|\xi_{i}\right\rangle $,
the final prepared mixed state will be $\hat{\tilde{\rho}}=\sum_{i}p_{i}\left|\tilde{\Psi}_{i}\right\rangle \left\langle \tilde{\Psi}_{i}\right|$.
If we assume that $\left\langle \left.\tilde{\Psi}_{i}\right|\Psi_{j}\right\rangle =0$
for $i\ne j$, then $\hat{\tilde{\rho}}$ suffers an error \citep{nielsen_quantum_2000}
\begin{eqnarray*}
\varepsilon_{\rho} & = & 1-\mathrm{Tr}\left(\sqrt{\hat{\tilde{\rho}}}\hat{\rho}\sqrt{\hat{\tilde{\rho}}}\right)^{1/2}\\
 & = & 1-\mathrm{Tr}\left(\sum_{i}p_{i}^{2}\left|\tilde{\Psi}_{i}\right\rangle \left\langle \left.\tilde{\Psi}_{i}\right|\Psi_{i}\right\rangle \left\langle \Psi_{i}\left|\tilde{\Psi}_{i}\right.\right\rangle \left\langle \tilde{\Psi}_{i}\right|\right)^{1/2}\\
 & = & 1-\sum_{i}p_{i}(1-\varepsilon_{\Psi,i})\\
 & \le & \varepsilon_{\Psi}\leq ml\varepsilon_{I}/2,\end{eqnarray*}
where $\varepsilon_{\Psi}=\max\varepsilon_{\Psi,i}$. That is, the
assembly of mixed states does not magnify the pre-existing errors.

Overall, we see that errors introduced in any stage of the state preparation
algorithm are not amplified more than polynomially by subsequent stages.
The final error in the prepared state is $\varepsilon=\varepsilon_{\rho}\leq ml\varepsilon_{I}/2$,
meaning that the error scales linearly with the size of the system
$m$ and the error of the integration procedure, as well as logarithmically
with the grid size $2^{l}$.

\subsection{Computational cost}

\label{sub:scaling}

There are three time-consuming steps in the state preparation algorithm.
The first is the evaluation of the integrals \eqref{eq:integrals}
and the resulting single-qubit rotations, the second is the phase-estimation
that is used to distinguish states in the superposition (see Sec.
\ref{sub:superpositions}), and the final is the antisymmetrization
procedure described in Sec. \ref{sub:many}. We characterize the cost
of each step in turn.

In the previous section, we have seen that the total error of the
prepared state will be $\varepsilon\leq ml\varepsilon_{I}/2$. Therefore,
if we want to ensure a maximum error $\varepsilon$, we must choose
$\varepsilon_{I}=2\varepsilon/ml$, implying that $O(ml\varepsilon^{-1})$
time is required for each integration (see Sec. \ref{sub:efficiency}).
The integration procedure itself is called $ml$ times: for each of
the $m$ occupied orbitals, $l$ qubits have to be rotated correctly.
Therefore, the total time required for all the qubit rotations is
$O(m^{2}l^{2}\varepsilon^{-1})$.

The cost of the phase-estimation procedure that is used to distinguish
the eigenstates cannot be given precisely because we have not made
any assumptions about the nature of the Fock operator $\hat{F}$ other
than that it is efficiently simulatable, that is, running in time $\mathrm{poly}(m,M,l,\Delta^{-1})$
(here, $\Delta$ is the precision at which the simulation needs to be
run, i.e., it is half the gap between the closest two eigenstates,
which we assumed is not exponentially small).
Simulating the entire Hartree-Fock Hamiltonian requires the simulation
of the Fock operator acting separately on each particle, meaning that
the total simulation requires $m\mathrm{poly}(m,M,l,\Delta^{-1})$ time. In addition
to this, two quantum Fourier transforms (QFTs) are required on the
readout register of the phase estimation. If $q$ qubits are used
for the readout (see Sec. \ref{sub:errors}.4), the QFTs require $O(q^{2})$
time. It should be noted that the required $q$ is determined only
by needed precision in the phase estimation, and that it does not
depend strongly on $m$, $l$, or $M$. Therefore, the cost of the
QFTs can be treated as essentially a constant overhead. Furthermore,
there is the cost of looking up the state's energy in the look-up
table; a simple binary search requires $O(\log_{2}M)$ time per register,
for a total cost of $O(m\log_{2}M)$. But this, too, is a negligible
cost in comparison to $m\mathrm{poly}(m,M,l,\Delta^{-1})$, which we conclude is
the asymptotic cost of the eigenstate discrimination portion of the
state preparation algorithm.

The bottleneck of the antisymmetrization procedure used to produce
fermionic states (or the symmetrization for bosonic ones) is the sort
that takes state \eqref{eq:sort_before} to \eqref{eq:sort_after}.
Sorting register $B$ by a comparison sort requires $\Omega(m\log m)$
swaps. These swaps must also be performed on each of the corresponding
$l$ qubits of register $A$, for a total cost of $\Omega(lm\log m)$.
For large systems, this expression will be dominated by the scalings
of the integral evaluation and the phase estimation.

Based on the foregoing, the total computational cost of the state
preparation algorithm is
$O(m^{2}l^{2}\varepsilon^{-1}+m\mathrm{poly}(m,M,l,\Delta^{-1}))=\mathrm{poly}(m,M,l,\varepsilon^{-1},\Delta^{-1})$,
an expression polynomial in all the basic descriptors
of the system. This allows us to conclude that the algorithm, as described
above, is efficient.

\section{Conclusion}

\label{sec:conclusion}

We have outlined a quantum algorithm for the preparation of physically
realistic quantum states on a lattice. In particular, we have gone
beyond previous proposals by describing a method for preparing any
pure or mixed state of any number of particles. This is achieved by
using Zalka's method for preparing single-particle states and then
combining those into many-particle states. The assembly of many-particle
states requires that we be able to distinguish them on a quantum computer,
a task that we address using phase estimation. We also provided symmetry-based
solutions for degenerate cases, where phase estimation using a single
operator is insufficient to distinguish the states. Accidentally
degenerate states can be distinguished by adding a perturbation to
the system Hamiltonian. Our algorithm is efficient,
with a run-time of $\mathrm{poly}(m,M,l,\varepsilon^{-1},\Delta^{-1})$,
subject only to the requirements that the wavefunction be bounded
or that its indefinite integral be known and that the Fock operator
be efficiently simulatable.

\begin{acknowledgments}
We acknowledge support from the Army Research Office under contract
W911NF-07-0304. NJW thanks the Harvard College Research Program and
IK the Joyce and Zlatko Balokovi{\'c} Scholarship.
\end{acknowledgments}

\end{document}